\documentclass{llncs}

\usepackage[T1]{fontenc}
\usepackage{graphicx}
\usepackage{booktabs}
\usepackage{array}
\usepackage{multirow}
\usepackage{enumitem}
\usepackage{cite}
\usepackage[hidelinks]{hyperref}
\setlist{nosep,leftmargin=*}
\institute{Indian Institute of Technology Ropar, Rupnagar, Punjab, India\\
\email{\{meenakshi.19csz0013,rohit.24csz0014,\\sakshi.23csz0006,sudarshan\}@iitrpr.ac.in}
\and
ANNAM.AI, Indian Institute of Technology Ropar, Rupnagar, Punjab, India\\
\email{\{pavania.harvard2025,adityabmv,jinalbirla,prakash.hegade\}@gmail.com}}

\begin{document}

\title{Effects of a Behavioural Commitment Scheme on Study Regularity in a Self-Paced
Learning Platform}
\titlerunning{A Behavioural Commitment Scheme in Self-Paced Learning}

\author{Meenakshi V\inst{1} \and Pavani Ayinampudi\inst{2} \and Aditya B.M.V.\inst{2} \and
Jinal Gupta\inst{2} \and Prakash Hegade\inst{2} \and Rohit Sharma\inst{1} \and
Sakshi Sharma\inst{1} \and S.R.S. Iyengar\inst{1}}
\authorrunning{Meenakshi V et al.}

\maketitle

\begin{abstract}
Online education has enabled learners worldwide to take up courses from reputed
institutions. However, a self-paced online course cannot
guarantee the motivation and engagement that a learner experiences in a real-time
classroom. Self-paced access also makes platform load unpredictable, which drives up the
compute cost.

\smallskip\noindent We propose a Commitment Scheme for course access that aligns learner
commitment with platform capacity. Learners book their study slots in advance; the
instructor sets the budget of learning hours available for the course; and learners who
use a full window earn additional watch hours. A booked window
records an intention to study at a stated time, and a learner who appears in that window
implements it. The byproduct is a platform load that can be forecast and bounded.

\smallskip\noindent The study involved two courses taken in sequence by the same learners, with the slot
booking system activated only in the second.
In-window study was observed on 86.8\% of booked windows, and the median committer placed
95.5\% of all study time inside self-booked windows. Among learners who studied across the
launch, study regularity improved from 1.01 to 1.33 active days per week, with a supporting
difference of +1.37 days per week against the same learners' preceding course. Commitments made on the same day as the study slot were
honoured more often than advance bookings (88.8\% against 62.5\% two days ahead), which is
consistent with the classic intention-behaviour gap.

\keywords{capacity planning \and commitment devices \and implementation intentions \and intention-behaviour gap \and self-regulated learning \and  slot booking }
\end{abstract}

\section{Introduction}

Online education now carries university instruction across geographical boundaries, with
national platforms such as NPTEL and SWAYAM serving millions of self-paced learners
\cite{kaveri2016}. The model rests on a quiet assumption: that learners left
to schedule themselves will actually study. A self-paced course offers none of the
classroom's scaffolding. There are no fixed hours, no visible peers, and no instructor who
notices absence. What the platform records is completion, and completion certifies clicks
rather than the disciplined, distributed study that learning science demands \cite{boroujeni2018}.

The failure is predictable rather than accidental, and hyperbolic discounting
\cite{laibson1997} explains why. A study plan made on Monday for Wednesday is sincere, but
when Wednesday arrives the immediate cost of studying outweighs the still-distant
certificate, and the plan is abandoned by the very person who made it. The result is the
intention-behaviour gap \cite{sheeran2002}, in which roughly one in three
to one in two intended actions never happen. Behavioural science offers counters:
implementation intentions, if-then plans of the form ``I will study in window X on day
Y'', reliably narrow the gap \cite{gollwitzer1999,gollwitzersheeran2006}, and commitment
devices bind the present self against the future self's defection \cite{ariely2002}. Deployed at scale, though, the record is uneven
(Section~2.1): scheduling offered as advice or as an optional tool helps unevenly or not at
all \cite{yeomans2017,patterson2018,kizilcec2017,baker2016,davis2016,blondeel2024,heikkinen2023}. What remains
untested is scheduling that is binding, verified, and the default way into the course.

Self-paced access also creates a second problem: load. When anyone may study at any
instant, demand is unforeseeable, and the standard answer of reactive autoscaling is the
costly one, since it lags demand and provisions for a population peak that rarely arrives
\cite{armbrust2010,loridobotran2014,qu2018}. Our own backend showed the textbook symptom
before this study began, saturating daily despite serving only a few hundred active
learners. Forecast-driven provisioning is the known remedy \cite{roy2011,maohumphrey2011},
but it needs a demand forecast that self-paced platforms, by construction, do not have.

We propose that the two problems share one solution. In our slot-booking commitment scheme,
deployed on ViBe, an AI-enabled learning platform, learners pre-book fixed two-hour study
windows. Entering the course requires a current booking, an instructor-set hours budget bounds
consumption, in-window activity is verified from watch telemetry, and a fully used window earns
a same-day bonus booking. Pedagogically, booking scaffolds the forethought phase of
self-regulated learning \cite{zimmerman2002,edisherashvili2022} and makes commitment the
default rather than an optional tool. Restricting course entry to the booked slot helps the
learner resist present bias, while the bonus incentivises adherence, engaging hyperbolic
discounting twice. Operationally, the booking log is an appointment book, which has long been an
instrument for observing show-up rates, no-shows, and over-booking in service-operations
research \cite{bailey1952,cayirliveral2003,laganga2007}. This booking log is now available
two days ahead as a load forecast for compute capacity.

We evaluate the scheme in a live 946-learner web-development course in a national summer internship
programme, where it launched 37 days in and ran for 25 observed study days (1,862 evaluable
bookings by 317 learners). The mid-course launch is both a threat and an instrument, since it
splits learners by exposure (Section~4.2) and yields a within-learner interrupted time series
plus a difference-in-differences comparison against a second, unscheduled course that the
same learners take before this one. Adoption is self-selected, so the design is quasi-experimental, and we scope
every claim accordingly.

The paper contributes: (1) the first system, to our knowledge, in which one learner action,
booking a study window, is both a verified self-regulated-learning commitment and the
platform's capacity-planning input; (2) evidence that binding scheduling can succeed where
nudged scheduling failed, with 86.8\% of 1,862 bookings honoured and study regularity among
committed learners improving from 1.01 to 1.33 active days per week, supported by a +1.37
difference-in-differences against their preceding course; (3) a measured load forecast, in which booked
seats explained 59\% of the variance in actual concurrent load and bounded its peak, though
the cost saving itself remains a projection; and (4) a methodological lesson, namely that
mid-course launches demand exposure-aware cohorting. While the lead-time gradient is consistent
with established intention-behaviour theory, this study extends it in three respects: the
gradient is observed in behavioural telemetry rather than self-report, it arises under a binding
rather than a nudged design, and it serves as the decay factor that converts a commitment log
into a capacity forecast.

\section{Related Work}

\subsection{The Intention-Behaviour Gap, Implementation Intentions, and MOOC Scheduling}

Intentions are weak predictors of behaviour: deciding to act explains under a third of the
variance in acting \cite{sheeran2002}. Hyperbolic discounting \cite{laibson1997} formalises the
dynamic inconsistency by which future plans are reversed as execution nears, and
construal-level theory adds that distant events are
represented abstractly and near events concretely \cite{tropeliberman2010}, so a two-day-ahead
commitment is a vague good intention, compounded by planning-fallacy optimism about one's
future self \cite{buehler1994}. Implementation intentions, which pre-decide when and where and
convert a goal into a situational trigger \cite{gollwitzer1999,gollwitzersheeran2006}, are the
canonical counter, with direct evidence for study time specifically \cite{duckworth2013,felker2023}. At MOOC scale, these ideas form a clear progression. Writing a
concrete study plan raised completion by roughly 29\% relative \cite{yeomans2017}, reserving
weekly study time predicts goal attainment \cite{kizilcec2017}, a self-imposed daily
time-limit device raised study time 24\% and completion 40\% \cite{patterson2018}, and daily
micro-commitments reduced procrastination in online courses \cite{felkey2023}. But the
intervention structurally closest to ours is a null result: Baker, Evans and Dee
\cite{baker2016} randomised 18,043 learners into an invitation to pre-schedule lecture time,
with no improvement in persistence, and voluntary SRL scaffolds at scale are simply not taken
up \cite{davis2016}. The gap is that no prior system made scheduling binding (the access
path), verified (checked against telemetry), and default, which is the corner our deployment
occupies.

\subsection{Habit, Regularity, and Onset Timing}

Habits form through repetition in stable contexts \cite{lally2010}, become context-cued rather
than goal-driven, and resist \emph{retrofit} onto an already-established routine
\cite{woodtamwitt2005,neal2011}, while new behaviour installs most easily at a temporal landmark
such as a course's start \cite{dai2014}. Study-time \emph{regularity}, meaning consistency as
distinct from volume, is a measurable, outcome-linked construct in MOOC sequences
\cite{boroujeni2018,li2022}, and it is accordingly our primary outcome.

\subsection{The Operations Bridge: Appointments and Cloud Capacity}

A learner who books and does not appear is, formally, a clinic patient missing an appointment,
the subject of seventy years of operations research from Bailey's founding queueing analysis
\cite{bailey1952} through canonical reviews \cite{cayirliveral2003,dantas2018}. Average clinic
no-show is around 23\% against our 13.2\%, and lead time is the dominant no-show driver
\cite{dantas2018,gallucci2005}, which is the shape we observe. Providers overbook against
expected show-up probability \cite{laganga2007,zachariaspinedo2014}. Our booking log is
literally this artefact, and it adds a learning-science account of \emph{why} lead time predicts
no-show. On the systems side, the elasticity literature diagnoses reactive autoscaling as the
weakest strategy and names workload-prediction accuracy as the bottleneck of proactive
alternatives \cite{loridobotran2014,qu2018}, while known future demand cuts provisioning cost
substantially \cite{roy2011,maohumphrey2011}. Over-provisioning for peak is the core economic problem
cloud elasticity was built to solve \cite{armbrust2010}. A reservation log sidesteps the
prediction bottleneck entirely, because demand \emph{declares itself}.

\section{The Slot-Booking Commitment Scheme}

\subsection{Platform Context}

ViBe is a MERN-stack (MongoDB, Express, React, Node) online learning platform used in a national
summer internship programme. Courses consist of videos with in-video checkpoints, quizzes, blogs, and
projects. The player emits a watch-telemetry record for roughly every 15 seconds of active
playback, capturing who, which item, start, and end. This stream, some 400,000 sessions in the
studied course, is the paper's behavioural instrument. The backend runs on serverless cloud
infrastructure (Google Cloud Run) whose default posture is reactive autoscaling.

\subsection{Mechanics and Their Rationale}

The scheme is built from five pieces that hand off to one another. A learner books a
window, the booking unlocks the course, the player's telemetry records what actually
happened inside the window, an evaluator settles whether the commitment was kept, and kept
commitments earn more time.

\begin{itemize}
\item \textbf{Booking.} Every day offers the same seven two-hour windows, from 10:00 to just
before midnight IST. A learner can reserve a window for study day $D$ at any time between
09:00 on $D-2$ and 09:00 on $D$ itself, and receives one base booking per calendar day. A
booking is a small promise with a timestamp, recorded before the moment arrives.

\item \textbf{Access gate.} The course opens only for a learner whose booked window is
currently running. We gate entry alone, so a learner who begins inside their window may keep
going after it ends: the design governs when study \emph{begins}, not when it stops. Where
the gate sits is what makes it work. It stands where a present-biased learner would otherwise
defect, and because booking is the only way in, the commitment is not an optional planner
that a struggling learner can quietly ignore.

\item \textbf{Hours budget.} Booking draws on a finite account of bookable hours
that the instructor allots for the course, 21 at the time of the experiment, sized from the
estimated effort of the content. Every reservation spends its two hours against that budget,
and the instructor can top up individual learners where needed.

\item \textbf{Verified fulfilment.} When a window closes, an evaluator reads the watch
telemetry and totals how long the learner was active inside it. At the time of the study, a
window counted as fulfilled only if the learner had been active for at least 90\% of it.

\item \textbf{Bonus loop.} Keeping a commitment pays out immediately, because a fulfilled
window grants one extra booking usable the same day. Where the gate works \emph{against}
present bias, the bonus puts it to work: the reward for studying now is more access now.
\end{itemize}

\subsection{The Byproduct, by Construction}

The bookings table, holding learner, study date, and window, is a per-window demand schedule
known up to two days in advance. Summed per cell, it tells the provider how many learners have
declared they will study in each window of each day: an appointment book for compute capacity,
evaluated as a forecast in Section~5.5.

\section{Study Design and Method}

\subsection{Setting and Scale}

The studied course is a MERN web-development course in the 2026 edition of a national summer internship
programme. It opened on 21 May 2026, the commitment scheme launched on 27 June, and our
observation window closes at the end of 21 July, giving 25 study days of scheme operation. The
dataset was then frozen, so all numbers come from a single locked extraction. There were 946
enrolled learners (919 with any watch activity) and 1,875 bookings, of which \textbf{1,862} are
evaluable, made by \textbf{317} distinct committers (34\% of active enrolment) at a median of 6
bookings each.

\subsection{Exposure Cohorts: The Mid-Course Launch as Instrument}

A feature launched mid-course cannot be evaluated by pooled before-and-after averages, because
learners who finished before the launch contribute zeros afterwards \emph{mechanically}, which
manufactures artefacts in whichever direction one pools. We therefore partition all learners by
what they could have experienced (Table~1). \textbf{C1} were active only before launch and
never booked, so they provide baseline behaviour and are excluded from all effect estimates.
\textbf{C2} were active across the launch and adopted booking, so each is their own control in a
within-learner interrupted time series, the study's primary evidence. \textbf{C3} first appeared
on or after launch, so booked study is the only course-access model they ever knew, making them
the robustness cohort. We never compare completion across cohorts, which differ in
time-in-course.

\begin{table}[tb]
\caption{Deployment summary and exposure cohorts (C1 baseline, C2 primary, C3 robustness)}
\centering
\footnotesize
\begin{tabular}{@{}lc@{\hspace{6pt}}l p{2.4cm} c@{}}
\toprule
\multicolumn{2}{c}{\textbf{Deployment}} & \multicolumn{3}{c}{\textbf{Exposure cohorts}} \\
\cmidrule(r){1-2}\cmidrule(l){3-5}
Enrolment (with activity) & 946 (919) & C1 & Pre-launch only; never booked & 603 \\
Scheme live & 25 study days & C2 & Across launch; adopted booking & 180 \\
Bookable windows/day & 7 $\times$ 2h & C3 & On/after launch; all booked & 136 \\
Evaluable bookings & 1,862 (317) & -- & Enrolled, never watched & 27 \\
Bookings/learner, med (max) & 6 (16) & & & \\
Booked day$\times$window cells & 172/175 & & & \\
\bottomrule
\end{tabular}
\end{table}

\subsection{The Control Course}

The same programme ran a course, Fundamentals of AI on the same platform with
\textbf{no} booking scheme. The programme follows a linear, sequential path with a common
offering structure: a learner who completes the AI course is then invited to the MERN course,
where the slot-booking scheme is enforced. This gives a within-person comparison, since 931
of our learners were enrolled in both, and the same individuals are observed with and without
the scheme. We use it for a difference-in-differences (DiD) estimate over the same calendar
weeks: for each learner, the change in the committed course minus the change in the control
course, which nets out stable personal traits. Because the courses are sequential, the control
course was 97.7\% complete when the scheme launched, so its activity over these weeks mainly
reflects completion. We therefore read the DiD as supporting evidence and treat the
within-learner change as the primary estimate.

\subsection{Measures}

All behavioural measures are computed by overlapping raw watch-telemetry sessions against
booked windows. The platform's stored fulfilment labels were contaminated by a since-fixed
evaluator bug and are not used anywhere.
\begin{itemize}[nosep]
\item \textbf{Per booked window.} \textbf{Show-up}: more than 1 minute of in-window activity;
\textbf{sustained use}: active for at least half the window, about 1 hour; \textbf{active
share}: fraction of the window active.
\item \textbf{Per learner.} \textbf{In-window share}: fraction of their study time falling
inside their own booked windows; \textbf{regularity}: active study-days per week. Regularity
is our primary outcome, chosen over hour-sums because a per-day binary is immune to the
double-counting that overlapping telemetry sessions inflict on duration sums, and because
day-spacing is the behavioural signature of distributed practice.
\item \textbf{Per booking.} \textbf{Lead time}: days between making the booking and the study
day.
\item \textbf{Per day$\times$window cell.} \textbf{Actual concurrency}: peak simultaneous watch
sessions, by event-sweep.
\end{itemize}
Because telemetry captures only video-player activity, all in-window figures are lower bounds.

\subsection{Ethics}

All analysis used de-identified, pseudonymised telemetry (salted hashes, no personally
identifying information in any export), read-only against the production database, under the
programme's data-use provisions.

\section{Results}

We report five results, in which effect estimates draw only on the committed
cohorts (C2, C3) and the rules of Section~4 apply throughout.

\subsection{R1: Learners Kept the Commitments They Made}

Across the full census of 1,862 booked windows (Table~2), \textbf{86.8\%} received in-window
study and \textbf{54.9\%} received a sustained session of at least half the window (median active
share 54.4\%). Commitment also organised \emph{when} study occurred, since the median committer
placed \textbf{95.5\%} of all study time inside self-booked windows. Booking was therefore the
structure within which study happened, not an annotation on study that would have happened
anyway. Both figures take intentions from the booking store and behaviour from watch telemetry,
over the census of commitments rather than a sample.

\begin{table}[tb]
\caption{Commitment follow-through (n = 1,862 booked windows)}
\centering
\small
\begin{tabular}{lc}
\toprule
Measure & Value \\
\midrule
Showed up (>1 min in-window) & \textbf{86.8\%} \\
Sustained session ($\geq \frac{1}{2}$ window) & \textbf{54.9\%} \\
Active share of window, median (mean) & 54.4\% (47.7\%) \\
Study time inside own windows, median (mean; n=294) & \textbf{95.5\%} (76.6\%) \\
\bottomrule
\end{tabular}
\end{table}

\subsection{R2: Study Regularity Improved among Committed Learners}

Table~3 and Fig.~1a present the primary outcome in three views. Within
learners, the 180 C2 students rose from \textbf{1.01 to 1.33} active days per week at booking
onset (median change +0.17), and 75\% of all bookers increased. The same learners'
preceding course offers a supporting view. The per-learner DiD is \textbf{+1.37} active days
per week (median +1.22): regularity rose in the committed course while activity in the control
course tapered as learners completed it (Section~4.3), so we take the within-learner change as
the primary estimate. Finally, C3 learners, who never experienced the course without commitment,
sustained \textbf{1.54} active days per week from their first week.

However, across all 810 shared learners the DiD is approximately zero
(+0.04, median $-$0.19), because most had completed the course before launch. The claim is that
commitment regularised study \emph{among the learners who adopted it}, a quasi-experimental
within-learner result over 317 self-selected committers, not a population or randomised
effect.

\begin{table}[tb]
\caption{Study regularity (active study-days per week)}
\centering
\small
\begin{tabular}{lcccc}
\toprule
View & n & Pre & Post & Effect \\
\midrule
C2 within-learner (own control) & 180 & 1.01 & 1.33 & \textbf{+0.32} (median +0.17) \\
Bookers who increased & 312 & -- & -- & \textbf{75\%} (234/312) \\
DiD, committed learners$^\dagger$ & 247 & -- & -- & \textbf{+1.37} (median +1.22) \\
DiD, all shared learners$^\dagger$ & 810 & -- & -- & +0.04 (median $-$0.19) $\approx$ 0 \\
C3 from-start & 136 & 0 & 1.54 & median $\Delta$ +1.68 \\
\bottomrule
\end{tabular}

\smallskip\footnotesize $^\dagger$Against the preceding AI course, taken
sequentially by the same learners (Sec.~4.3); supporting view.
\end{table}

\subsection{R3: Follow-Through Decayed with the Distance between Intention and Action}

Learners overwhelmingly committed at the point of intention, since \textbf{87.5\%} of
bookings were made on the study day itself, and follow-through decayed monotonically with
lead time (Table~4, Fig.~1b): 88.8\% same-day, 73.6\% one day ahead, 62.5\% two days ahead,
a significant downward trend across lead times (Cochran--Armitage trend test, $z = -6.8$,
$p < 0.001$; booking-level).
This is the intention-behaviour gap of Section~2.1, observed in server telemetry rather than
in self-report, and it matches what construal-level theory predicts: near plans are concrete
and are kept, while distant plans are abstract and are displaced.

\begin{table}[tb]
\caption{Follow-through by booking lead time}
\centering
\small
\begin{tabular}{lccc}
\toprule
Booked & n (share) & Showed up & Mean active share \\
\midrule
Same day & 1,630 (87.5\%) & \textbf{88.8\%} & 49.0\% \\
1 day ahead & 216 (11.6\%) & 73.6\% & 38.8\% \\
2 days ahead & 16 (0.9\%) & 62.5\% & 34.9\% \\
\bottomrule
\end{tabular}
\end{table}

\begin{figure}[tb]
\centering
\includegraphics[width=0.43\textwidth]{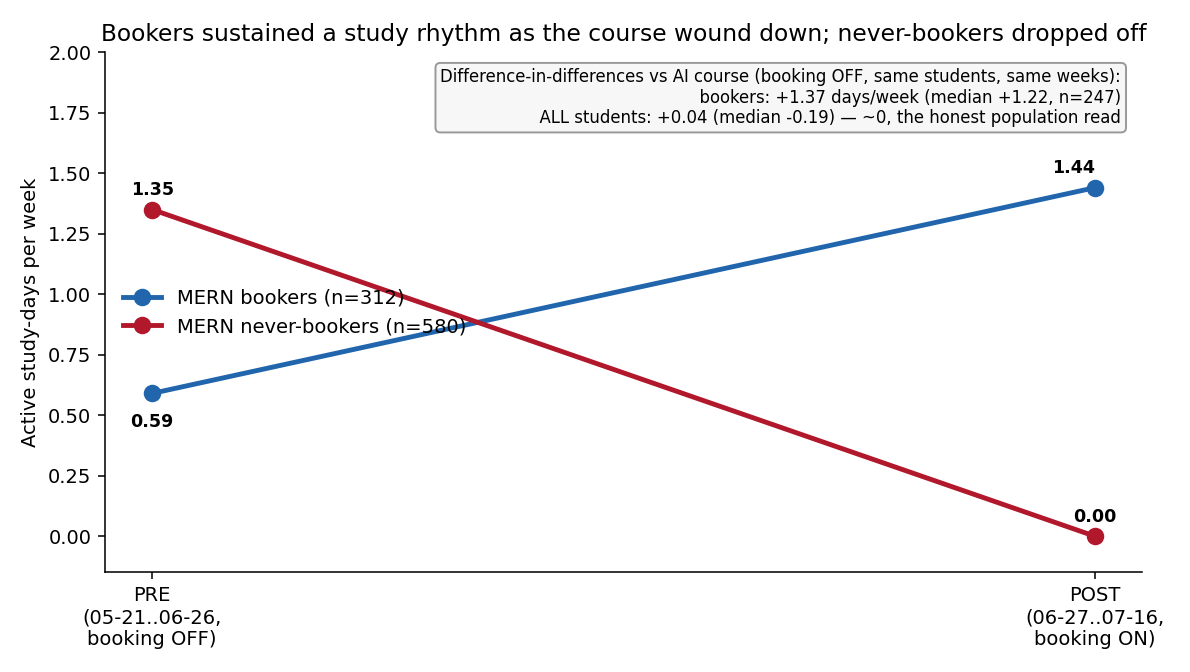}\hfill
\includegraphics[width=0.43\textwidth]{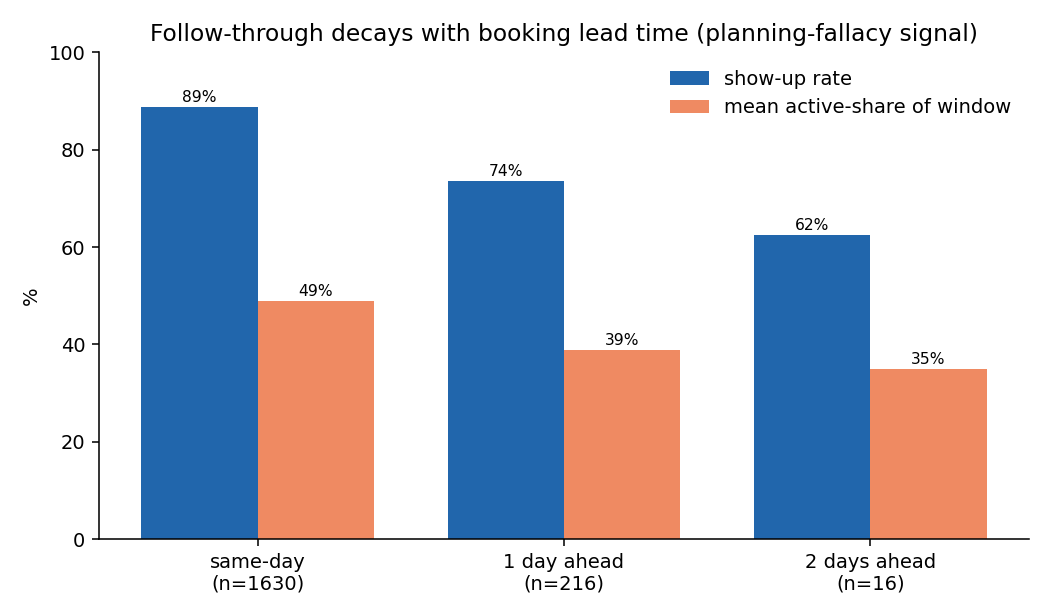}
\caption{(a) Study regularity (active days/week) before and after booking onset, by cohort
view. (b) Follow-through (show-up rate) decaying with booking lead time.}
\end{figure}

\subsection{R4: Commitment from the Outset Outperformed Mid-Course Adoption}

On volitional measures, meaning behaviour within windows each cohort itself chose to book,
from-start learners (C3) exceeded mid-course adopters (C2) on show-up (\textbf{92.1\% vs
82.5\%}) and mean active share (\textbf{55.0\% vs 41.6\%}), as Table~5 shows. Learners for whom
booked study was the only access model they had ever known adhered more completely than learners
onto whose five weeks of free-study habit the scheme was retrofitted, consistent with commitment
operating best as a default. The comparison is observational, since C3 learners were earlier in
their journey and self-selected by later enrolment. C3's in-window concentration (98.6\%) is
reported for completeness but excluded from the comparison, being partly imposed by the gate.

\begin{table}[tb]
\caption{Volitional adherence: mid-course adopters vs from-start committers}
\centering
\small
\begin{tabular}{lcc}
\toprule
Measure & C2 (n=180) & C3 (n=136) \\
\midrule
Show-up on own booked windows & 82.5\% & \textbf{92.1\%} \\
Mean active share of window & 41.6\% & \textbf{55.0\%} \\
Same-day booking share & 87.9\% & 87.3\% \\
\emph{(Context, not compared: median in-window study share)} & \emph{71.4\%} & \emph{98.6\%} \\
\bottomrule
\end{tabular}
\end{table}

\subsection{R5: The Booking Schedule Was a Measured Forecast of Actual Load}

Table~6 and Fig.~2 evaluate the operational byproduct. Correlating booked seats against actual
concurrent watch sessions per day$\times$window cell gives \textbf{r = 0.77
(R\textsuperscript{2} = 0.59)} for within-window peak concurrency, with consistent values for
distinct active learners (0.72) and delivered load-hours (0.75). The schedule was
\emph{bounding}, since the deployment-wide peak of actual concurrent sessions (25) stayed within
the booked-seat peak (28), and \emph{nearly complete}: \textbf{88.9\%} of delivered load fell in
forecast-visible cells, the remainder being overnight continuation beyond the bookable span
(10.0\%) and activity in unbooked windows (1.1\%). Booked demand ran an order of magnitude below
population, at a peak of 28 seats against 317 booking learners.

This result establishes a measured, up-to-two-days-ahead
forecast of course-scoped load, but \textbf{not} a measured infrastructure saving. The backend ran
fixed-size throughout (35.3 vs 35.7 instance-hours/day, pre vs post) because provisioning was
never coupled to the schedule. Service-level cloud metrics, aggregated and supply-saturated
in both periods, resolve none of this structure (per-window R\textsuperscript{2} $\leq$
0.06), confirming course-scoped telemetry as the correct instrument.

\begin{table}[tb]
\caption{Forecast quality: booked demand vs actual load (175 day$\times$window cells)}
\centering
\footnotesize
\begin{tabular}{lll}
\toprule
Property & Measure & Value \\
\midrule
Accuracy & booked seats vs peak concurrency & \textbf{r = 0.77 (R\textsuperscript{2} = 0.59)} \\
 & vs active learners / load-hours & r = 0.72 / 0.75 \\
Boundedness & actual peak vs booked peak & \textbf{25 $\leq$ 28} \\
Completeness & load in forecast-visible cells & \textbf{88.9\%} \\
 & outside bookable span / unbooked cells & 10.0\% / 1.1\% \\
Concentration & peak booked seats vs bookers & 28 vs 317 ($\sim$11$\times$) \\
Horizon & schedule known ahead & up to 2 days \\
Capacity (context) & instance-hours/day, pre/post & 35.3 $\rightarrow$ 35.7 \\
\bottomrule
\end{tabular}
\end{table}

\begin{figure}[tb]
\centering
\includegraphics[width=0.42\textwidth]{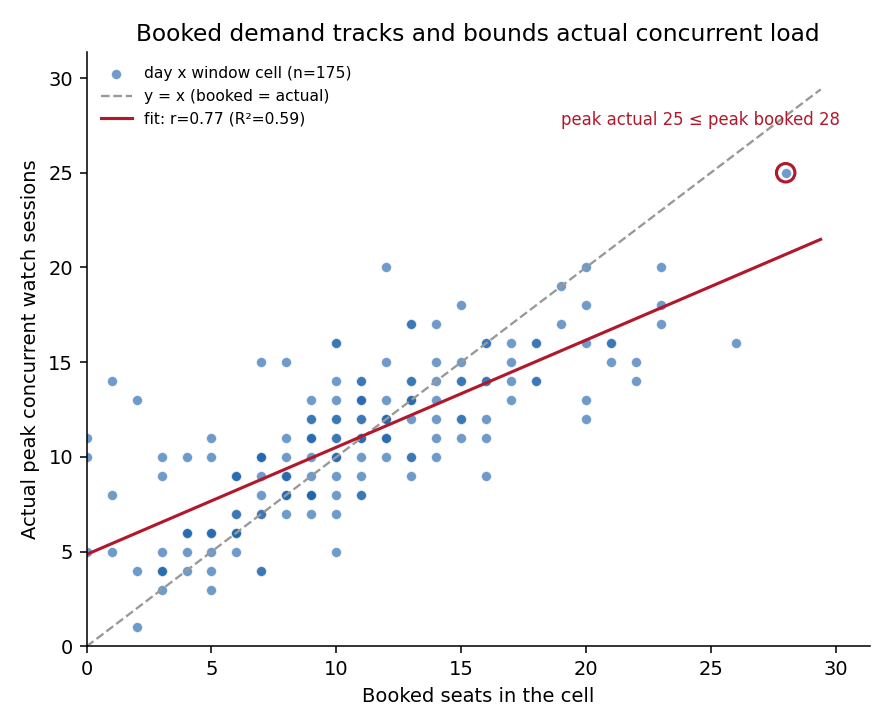}
\caption{Booked seats vs actual peak concurrent load, per day$\times$window cell (n=175). The
fit line gives the measured r/R\textsuperscript{2}; the marked point is the deployment-wide
peak, showing actual load (25) bounded by booked seats (28).}
\end{figure}

\section{Discussion}

\subsection{Why Binding Scheduling Succeeded Where Nudged Scheduling Failed}

The nearest precedent is the null result of \cite{baker2016} (Section~2.1), and three
properties explain the divergence. \textbf{Binding}: entry was gated on a current booking,
pre-commitment in the strict hyperbolic-discounting sense \cite{laibson1997}, closing exactly
the defection an informational nudge leaves available. \textbf{Verified}: fulfilment was
measured from telemetry, not claimed, the accountability ingredient separating a commitment
device \cite{patterson2018} from a reminder. \textbf{Default}: where voluntary SRL scaffolds
see chronically low uptake
\cite{davis2016}, a gate every learner passes through has complete uptake by construction. The
active ingredient is therefore not \emph{scheduling} as such but commitment made structural.
However, the three properties operated together. Hence, their separate roles in reducing
procrastination, building accountability, and forming habit cannot be isolated in this design.
The comparison remains interpretive rather than a controlled head-to-head, since Baker et al.\
randomised an unenforced nudge within one course while ours is a quasi-experimental deployment
in a different setting; a randomised binding-versus-nudge trial is the natural next
experiment.

\subsection{Timing, Habit, and Who the Scheme Reached}

Learners did not use booking to plan ahead. Follow-through decayed monotonically with lead
time, construal theory's signature \cite{tropeliberman2010,buehler1994}, and booking
functioned as a start-of-study ritual, formed and executed almost immediately, exactly where
implementation intentions are strongest \cite{gollwitzerbrandstatter1997}. Platforms should
therefore optimise the same-day path rather than push advance planning that future selves
routinely break, and should treat advance bookings as a capacity signal with a known decay
factor. Descriptively, the most-booked window was late-night (22:00), yet actual study centred
near 16:20 IST: learners \emph{book} aspirationally and \emph{study} practically, so day-of
re-commitment prompts and revealed-behaviour slot suggestions are the corresponding, untested,
levers. Separately, from-start learners (C3) out-adhered mid-course adopters (C2) on every
volitional measure,
matching habit theory's prediction that habits are context-cued and resist retrofit
\cite{woodtamwitt2005,neal2011}, while a course's start uniquely favours new behaviour
\cite{dai2014}. For C3, booking was the only access model they ever knew, while C2 retained
around 30\% of study outside windows, the residue of five pre-scheme weeks. This onset-timing
finding is hypothesis-grade, since C3 were also earlier in their journey, self-selected, and
partly gate-confined, with a matched-course-position comparison as the decisive follow-up.
Suggestively, mid-course adopters entered below the non-adopting population's baseline (1.01 vs
1.29 active days/week) and ended above it, an equity-relevant direction rather than a claim.

\subsection{The Bonus Loop, Autonomy, and the Ethics of Binding Design}

The reinforcement loop behaved as designed. 28 learners earned 46 bonus windows, which
out-performed base bookings on every measure (show-up 91.3\% vs 86.7\%; active share 58.9\% vs
47.4\%). We report this descriptively, since bonus windows belong by construction to
already-engaged learners and contingent rewards carry a crowding-out risk. The deeper question a
binding scheme must answer is \textbf{autonomy}. Gating access is deliberately paternalistic,
and self-determination theory warns that controlling designs can undermine intrinsic motivation
\cite{deciryan2000}. Three things soften the concern: learners chose freely
within the structure (window, day, rebooking); commitment was self-imposed daily rather than
imposed once, so 87.5\% same-day booking reads as ritual rather than strain; and the gate is
entry-only, constraining when study begins but not its content, pace, or volume, which preserves
the flexibility central to self-paced learning. Windows spanning 10:00 to midnight and instructor top-ups offer some
flexibility to learners with fixed schedules. Whether the scheme is fair to such learners, and
how it affects motivation beyond 25 days, remain open questions that require learner-voice data.

\subsection{The Booking Log as an Appointment Book: Operations Implications}

Our show-up rate sits above the roughly 77\% average that clinic no-show reviews report, with
lead time the same dominant driver \cite{dantas2018,gallucci2005} (Section~2.3). The
overbooking tradition \cite{laganga2007,zachariaspinedo2014} gives the provisioning rule:
effective load $\approx$ booked seats $\times$ show-up rate $\times$ active share $\approx$
0.5--0.6$\times$ booked seats, refined by the measured envelope (peak 25 within 28) and the
quantified residual (11.1\% invisible to the forecast, mostly overnight spill-over). Against
the elasticity literature's diagnosis that workload-prediction accuracy bottlenecks proactive
autoscaling \cite{qu2018}, our answer is structural: demand is declared up to two days
ahead rather than predicted. At the scale of national platforms serving millions of
self-paced learners, provisioning to declared demand rather than to population peak would
let infrastructure scale with actual study rather than with enrolment. However, the saving
is a \textbf{projection}, because provisioning was never coupled to the
schedule, and realising the headroom is engineering future work. In addition, \emph{service-level} cloud
metrics, being aggregated and supply-capped, resolved none of this structure: course-scoped
telemetry was the instrument that worked, and other demand-shaping evaluations in ed-tech should
expect the same.

\subsection{Limitations and Threats to Validity}

These results should be read with their limits in view. The study is quasi-experimental:
learners chose to adopt the scheme, and it arrived mid-course, so while the within-learner
series and the same-learner control course rule out stable traits, they cannot rule out
time-varying self-selection or course-position effects. The lead-time test treats bookings as
independent, and student-level models of follow-through and regularity change are a natural
next step with the extended dataset. The effect claim belongs to the 317 committers, not
the whole cohort, since the population-level difference is essentially zero and the control
course was already 97.7\% complete at launch. Our telemetry sees only the video player, so
in-window figures are lower bounds, which is why we lean on the per-day regularity metric, being
immune to the double-counting that inflates hour totals. The from-start advantage remains
hypothesis-grade, since course position, enrolment timing, and the gate all covary with cohort.
Everything rests on one committed course, one programme, and 25 days. Hence, the effects
remain to be tested in other course types, longer courses, and learner populations beyond a
single national summer internship programme in India. However, the lead-time gradient (R3)
and the load forecast (R5) are drawn from within the committed course and do not depend on
the cross-course comparison. The regularity effect shrank towards the course tail (1.56 to
1.33 across snapshots), and we report the conservative frozen value.
Above all, the study measures regularity and in-window engagement, not completion or learning
outcomes. Regularity is an established correlate of achievement in MOOC sequences
\cite{boroujeni2018}, but whether booking-induced regularity translates into learning gain
remains the most important question for subsequent work.

\section{Conclusion}

A single learner action, booking a two-hour study window, served two purposes at once. For the
learner it was a verified commitment, held 86.8\% of the time; among committed learners, study
regularity improved from 1.01 to 1.33 active days per week, and a median 95.5\% of study fell
inside self-chosen windows. For the platform it was a demand schedule known up to two days ahead that explained
59\% of the variance in concurrent load and bounded its peak. Binding, verified, default
scheduling therefore succeeded where an unenforced nudge had not, and the same booking log that
disciplines study also makes capacity forecastable.

Our next steps follow from the limitations above. On the systems side, we will let the booking
schedule drive provisioning and measure the saving we have so far only projected. On the
learning side, the priority is connecting kept commitments to outcomes: does honouring more
windows translate into better quiz performance and learning gain? That is followed by a
randomised binding-versus-nudge comparison to put Section~6.1's interpretation on causal ground.
Above all, we want to hear from learners themselves about autonomy and fairness, and to try the
one lever the scheme entirely lacks: commitments visible to peers.

\bibliographystyle{splncs04-unsort}
\bibliography{references}

\end{document}